\documentclass[reprint,showpacs,showkeys,preprintnumbers,amsmath,amssymb,prb]{revtex4-1}
\usepackage[latin1]{inputenc}
\usepackage{graphicx}
\usepackage{dcolumn}
\usepackage{bm}

\begin{document}

\preprint{APS/123-QED}

\title{Mode Competition in Dual-Mode Quantum Dots Semiconductor Microlaser}

\author{Laurent Chusseau}%
\affiliation{Institut d'\'Electronique du Sud, UMR CNRS 5214, Université Montpellier 2, 34095 Montpellier, France}%
\email{chusseau@univ-montp2.fr}

\author{Fabrice Philippe}%
\affiliation{LIRMM, UMR CNRS 5506, 161 Rue Ada, 34392 Montpellier, France}

\author{Pierre Viktorovitch}%
\affiliation{Institut des Nanotechnologies de Lyon, UMR CNRS 5270, École Centrale de Lyon, 36 Avenue Guy de Collongue, 69134 Ecully, France}

\author{Xavier Letartre}%
\affiliation{Institut des Nanotechnologies de Lyon, UMR CNRS 5270, École Centrale de Lyon, 36 Avenue Guy de Collongue, 69134 Ecully, France}

\date{\today}

\begin{abstract}
This paper describes the modeling of quantum dots lasers with the aim of assessing the conditions for stable cw dual-mode operation when the mode separation lies in the THz range.
Several possible models suited for InAs quantum dots in InP barriers are analytically evaluated, in particular quantum dots electrically coupled through a direct exchange of excitation by the wetting layer or quantum dots optically coupled through the homogeneous broadening of their optical gain. A stable dual-mode regime is shown possible in all cases when quantum dots are used as active layer whereas a gain medium of quantum well or bulk type inevitably leads to bistable behavior.
The choice of a quantum dots gain medium perfectly matched the production of  dual-mode lasers devoted to THz generation by photomixing.
\end{abstract}


\maketitle

\section{Introduction}

CW THz and millimeter wave generation using the beating frequency of a dual-mode laser is a topic that has attracted  many efforts during recent years because there is currently a lack of versatile  and easy to use sources at these frequencies \cite{Tonouchi:2007}. To fill this gap, THz photomixing has many strengths, including the use of photonic technologies as the manufacture of laser diodes and complex optical systems or the ability to carry the beat on optical fiber. In all cases the conversion of the optical signal is provided by a photomixer which transforms the fast variations of the optical intensity in the electromagnetic radiation. Since a few years these components exist at once in GaAs technology for wavelengths $\approx 0.8 \, \mu$m \cite{McIntosh:1995} and InGaAs technology around $\approx 1.55 \, \mu$m \cite{Mangeney:2007}.

Regarding the optical beat source, several alternatives were considered. The easiest way is to combine two independent lasers that are stabilized in frequency and amplitude. This solution is now commercial but it is expensive because of the use of extended cavity lasers or high power DFB requiring sophisticated control electronics \cite{Wingender:2003}. In addition, the use of two independent lasers implies that the drift and noise are added in the beat noise which directly blames the linewidth of the THz signal generated. The implementation of this solution in the case of high-resolution spectroscopy requires a metrological stabilization of lasers that remains difficult in practice \cite{Mouret:2009}.

The use of a single laser seems more simple and should allow an improvement of compactness. Two solutions are possible. In the first the laser operates in pulsed mode either by gain-switching or by active or passive mode locking. The THz beating is then obtained by filtering the high-frequency harmonics of the pulse repetition. Because the consistency of the optical phase is maintained in these modes, extremely fine laser linewidths can be obtained \cite{Criado:2012a, Criado:2012}. Besides tunability weakness in the case of mode-locking, the major drawback of this technique lies in the sophisticated filtering process which is itself a source of high attenuation and that needs to be offset by several optical amplifiers. This undermines the simplicity and cost of final assembly. Alternatively the use of a suitable structure in the active layer allows for the dual-mode filtering  task within the passively mode-locked laser itself. Millimeter frequency generation has been demonstrated that way \cite {OBrien:2010}, however, the complexity is transferred to the laser fabrication that must include a specific longitudinal structure along the optical waveguide.

The second  solution to generate THz with a single component is to use a laser operating in the dual-mode   regime. Again several techniques have been proposed. First with the advent of manufacturing technologies,  semiconductor lasers can now include a complex structure with multiple DFB sections operating independently \cite{Chusseau:2006, Kim:2009}. The fact that the two modes inject each other can lead to instabilities and may seriously decrease the tuning range. An elegant solution has been demonstrated with a diode-pumped solid-state laser using either a separation of the two polarization eigenstates or a spatial separation of the two modes using an intracavity birefringent crystal  \cite{Brunel:1997}. These latter structures have shown excellent performance both in terms of noise and beating tunability  \cite{Czarny:2004, Baili:2009}.

To simplify, we may want to use a single Fabry-Perot edge-emitting laser. Under certain conditions, it is then possible to obtain structures exhibiting a dual-mode beating at the roundtrip time of the optical cavity \cite{Tani:2000, Latkowski:2008}. This was used once to build an very compact imaging system  at millimeter waves \cite{Shibuya:2007}. Similarly we may want to use a vertical-cavity semiconductor-laser which by construction has much less longitudinal modes available in the gain curve. One must then stack two cavities within the structure \cite{Brunner:2000a, Chusseau:2002a} to obtain the required two separate longitudinal modes spaced in frequency by an amount dependent of the optical coupling. This concept has been recently taken over by some of us with a single vertical-cavity Bragg-mirror resonator and a photonic-crystal supporting a slow optical Bloch mode \cite{Kusiaku:2011}. These structures have many advantages for  planar applications. Encapsulating the two cavities within the same small component ensures that any optical-cavity frequency-drift, for example due to temperature variations, will apply similarly to both modes. As a consequence the beating frequency will be insensitive to such drifts at first order. A built-in dual-mode laser is also an integrated component extremely compact and easily integrated into a system for which the stability of the beat will be easily obtained by monitoring the overall operating parameters of a single component (temperatures, currents, \dots). To enjoy these benefits, however, it is necessary that the modal competition does not destroy the desired dual-mode operation.

Modal competition is at the heart of the stable dual-mode operation of a single semiconductor laser. Since the seminal work of Lamb \cite{Sargent:1974}, it is well known that even if the laser cavity allows two simultaneous modes with similar quality factors and gains, the gain medium non-linearity sometimes forbids their simultaneous emission. A stable dual-mode operation is only allowed when the coupling factor between the modes is not strong enough. Otherwise the laser operates in a bistable regime where one mode dominates. Although unwanted,  this case unfortunately  happens with gain media such as bulk semiconductor or quantum wells (QWs). The physical origin is the short conduction-band intraband relaxation-time that couple two adjacent modes sharing the same carrier population. Following Agrawal \cite{Agrawal:1987} and taking an intraband relaxation time below 100~fs which is  consistent with InGaAsP active layers \cite{Yamada:1983} or InGaAs QW\cite{Debaiseux:1997}, it is not possible to have a stable dual-mode operation if the two modes are separated by less than 1.6~THz. This result was also obtained from rate-equations \cite{Chusseau:2003, Ahmed:2003}. Such an estimation moreover neglects the spatial hole burning which is also involved in the coupling of  longitudinal modes \cite{Lepage:2002}. This leaves no chance for a semiconductor laser emitting at $1.55 \, \mu$m and incorporating a bulk or  QW gain medium to operate dual-mode with a beating of $\approx 1$~THz. The  dynamics of multimode VCSEL \cite{Albert:2005} or edge-emitting laser \cite{Yacomotti:2004, Serrat:2006}, and even multisection DFB lasers \cite{Gao:2010} claims this.

Our dual-mode laser design involving a slow optical Bloch mode in a vertical cavity structure is supposed to operate dual-mode \cite{Kusiaku:2011}. It is thus mandatory to incorporate a gain medium that will not preclude the laser to operate in the desired regime because of a too strong coupling between modes. As shown by the above analysis, this coupling is mainly due to the sharing of the same carrier populations in the conduction band by the two adjacent modes. The proposed solution to circumvent this strong coupling is to use InAs quantum dots (QDs). The purpose of this work is then to verify that either optical or electrical coupling between InAs dot populations will remain at a sufficiently low level to maintain the dual-mode operation. Dots are optically coupled by the homogenous broadening of the optical gain that is estimated 
a few meV at 300~K \cite{Gammon:1996, Borri:1999, Matsuda:2001, Kammerer:2002, Bafna:2006}. Dots are also electrically coupled by their common interaction with the wetting layer that may allow a transfer of excitation from one dot to another.

The rest of the paper is organized as follows. Section \ref{unic} reminds the analytical modeling of dual-mode lasers with a single gain medium. Section \ref{QDmodel} details the case of dual-mode lasers including QDs by successively assuming first that dots are not coupled, and then coupled by a chemical-like equilibrium law linking excited dot populations through an interaction with the wetting layer, and finally assuming that the QDs are optically coupled through their homogeneous emission width. The conclusion is given in section \ref{conclusion}.

\section{Dual-mode semiconductor laser with a single gain medium}\label{unic}

A steady dual-mode operation requires special conditions for the allowed modes of a laser. The analysis initially given by Lamb \cite{Sargent:1974} is still the most relevant and easier to understand, it is based on two coupled differential equations that describe the time evolution of the modes in the laser cavity. These modes are characterized by their respective intensities $I_1$ and $I_2$\begin{subequations}
\begin{align}
\label{lamb}
\frac{d I_1}{d t}   & = \left( \alpha_1 - \beta_1 I_1 - \theta_{12} I_2 \right) I_1 \\
\frac{d I_2}{d t}   & = \left( \alpha_2 - \beta_2 I_2 - \theta_{21} I_1 \right) I_2
\end{align}
\end{subequations}
where  $\alpha_i$ are the unsaturated optical gains and where  $\beta_i$ and $\theta_{ij} $ are the  self-saturation and cross-saturation  gain coefficients. The stable solutions of these equations are obtained by canceling the derivatives
\begin{align}
\label{ssbi-mode}
I_1^\mathrm{ss}&= \frac{\alpha_1 - \left( \theta_{12} / \beta_2 \right) \alpha_2}{\left( 1 - C \right) \beta_1}  &
I_2^\mathrm{ss}&= \frac{\alpha_2 - \left( \theta_{21} / \beta_1 \right) \alpha_1}{\left( 1 - C \right) \beta_2}
\end{align}
with
\begin{equation*}
C= \frac{\theta_{12} \theta_{21}}{\beta_1 \beta_2}
\end{equation*}
a dimensionless coupling factor introduced by Lamb who showed using a perturbation analysis \cite{Sargent:1974} that it controls the existence and stability of the dual-mode operation. If $C> 1$, \emph{i.e.} in the case of a strong coupling between the modes, only single mode is enabled, the system is bistable and the laser operates continuously only on either of the two modes.

Starting from this conditions, it may be possible however to obtain a stable dual-mode regime changing the gain medium itself. This has been shown in several ways:
\begin{itemize}
\item by adding a saturable absorber in the cavity \cite{Chusseau:2006};
\item by physically decoupling the two gain areas  (for example by writing a dual-period DFB within the cavity \cite{Ginestar:2011,Lu:2013}, or a solid-state laser where the gain is physically separated into two different paths, for example by changing the polarization \cite{Czarny:2004}).
\end{itemize}

Neither of these solutions is satisfactory and feasible for our application where both modes share the same gain location \cite{Kusiaku:2011}. We will further investigate the link between this model and the most common rate-equations model of semiconductor lasers.

A semiconductor laser with a bulk or QW gain medium is generally described on the basis of population rate evolutions. Ahmed and Yamada \cite{Ahmed:2002} gave a fairly complete version of it from which one has extracted the equations describing a dual-mode laser
\begin{subequations}
\label{rate}
\begin{align}
\frac{d N}{d t} &= \kappa - \frac{N}{\tau_n} - A_1 N S_1 - A_2 N S_2  \\
\frac{d S_1}{d t} &= -\frac{S_1}{{\tau_p}_1} + A_1 N S_1 \left( 1 - \varepsilon_1 S_1 - \vartheta_1 S_2 \right) \\
\frac{d S_2}{d t} &= -\frac{S_2}{{\tau_p}_2} + A_2 N S_2 \left( 1 - \varepsilon_2 S_2 - \vartheta_2 S_1\right)
\end{align}
\end{subequations}
where $N$ is the carrier density beyond transparency, $\kappa = I / (e V)$ is the pumping parameter which depends directly on the injection current $I$ and volume of the active region $V$, $\tau_n$ is the carrier recombination time, the $A_i$ are the differential gains and $S_i$ the corresponding photon densities, ${\tau_p}_i$ are the photon lifetime within the cavity, $\varepsilon_i$ and $\vartheta_i$ the self-saturation and cross-saturation coefficients of optical gains. Indices $i$ refer of course to the various modes allowed by the cold cavity here reduced to 2. Considering a laser running well above threshold, it is possible to neglect the non-radiative recombination of carriers $N / \tau_n$ whose contribution can be seen as implicitly counted by a slight modification of the $\kappa$-values which reflects the pumping threshold.
In Eqs.~\eqref{rate} the coupling terms between the two modes appear explicitly with $\vartheta_i$. As their physical origin is similar to the self-saturation gain terms $\varepsilon_i$, these must also be taken into account. The omission of the correction factor $\left( 1 - \varepsilon_i S_i - \vartheta_i S_j \right)$ in the equation carriers however provides a great simplification of the calculations at the cost of a very weak approximation.

Like with the Lamb model, simple calculations allow to obtain the stationary solutions of Eqs.~\eqref{rate}. Among the solutions, only the one with simultaneous nonzero optical intensities $S_1$ and $S_2$ corresponds to a dual-mode laser, others accounting instead of bistable operations. In the general case these solutions are quite complicated but in the simplistic case where the  two modes are close to the gain  maximum  with a spacing small enough that their parameters are very similar, then $\varepsilon_i = \varepsilon$, $\vartheta_i = \vartheta$, ${\tau_p}_i = {\tau_p}$ and  $A_i = A$ and the stationary values are\label{steadystate}
\begin{align}
S_1^\mathrm{ss}=S_2^\mathrm{ss}& = \frac{\kappa}{2 u} &N^\mathrm{ss}&= \frac{u}{A} \label{bimodeSS}
\end{align}
with $u=\frac{1}{2}(\varepsilon+\vartheta)\kappa +\frac{1}{\tau_p}$.

The stability of this solution is further demonstrated by an analysis of variations around the operating point. To do this we replace in \eqref{rate} rates by their stationary values of \eqref{bimodeSS} plus a small variation, $N=N^\mathrm{ss}+n$, $S_i=S_i^\mathrm{ss}+s_i$. Keeping only the first order perturbations, we obtain a matrix system of three coupled first-order differential-equations that one writes in matrix form. The dual-mode stationary solution is then stable if the variations diminish over time, which is true if and only if the real parts of the eigenvalues of the corresponding matrix $\mathbf{\Theta}$ are negative.

In our particular case, the matrix is
\begin{equation}
\mathbf{\Theta} = \begin{pmatrix}
 -\frac{A \kappa}{u} &  -u & -u \\
 \frac{A \kappa}{2 u^2 \tau_p} &  -\frac{\varepsilon \kappa}{2} &   -\frac{\vartheta \kappa}{2} \\
 \frac{A \kappa}{2 u^2 \tau_p} &  -\frac{\vartheta \kappa}{2} &   -\frac{\varepsilon \kappa}{2}
\end{pmatrix}
\end{equation}
whose characteristic polynomial is
\begin{multline} \label{eigenvalues}
\mathcal{P}(\lambda)=- \left( \lambda - \frac{\kappa}{2} \left( \vartheta-\varepsilon \right) \right) \\
\left( 
\lambda^2 
+ \lambda \left( \frac{A \kappa}{u} + \frac{\kappa}{2} \left( \vartheta+\varepsilon \right) \right)
+ \frac{A \kappa}{u} \left( \frac{1}{\tau_p} + \frac{\kappa}{2} \left( \vartheta+\varepsilon \right)\right)
\right)
\end{multline}
The eigenvalues of $\mathbf{\Theta}$ are the roots of $\mathcal{P}$, that is $\frac{\kappa}{2} \left( \vartheta-\varepsilon \right)$ and the two roots of the polynomial of degree two on the right side. With our choice of parameters, all  coefficients of this polynomial are positive and then the two roots always exhibit negative real parts. Consequently, the dual-mode stationary solution proposed is stable if and only if $\vartheta < \varepsilon$.

The self-saturation and cross-saturation coefficients can be calculated using a quantum model of the gain medium and the result shows typically $\vartheta \approx 4 \varepsilon / 3$ \cite{Yamada:1986,Agrawal:1987,Ahmed:2002}. In such conditions, natural dual-mode behavior is not expected from a semiconductor laser, unless strong dispersive effects are included in the cavity to remove the condition assumed here: $\varepsilon_i = \varepsilon$, $\vartheta_i = \vartheta$, ${\tau_p}_i = {\tau_p}$ and  $A_i = A$. This was recently confirmed by numerical simulations of the dynamics of semiconductor lasers made with typical parameters from GaAs or InGaAsP technologies. Bistable behavior or highly multimode with a total power usually spread over a large number of modes is always obtained, in perfect agreement with experiments \cite{Ahmed:2002, Ahmed:2003}. This result was also confirmed with a Maxwellian modeling of laser  \cite{Serrat:2006} which only shows possibilities of either a single-mode behavior (actually bistable), or highly multimode. Both our analysis and these works indicate that dual-mode semiconductor lasers are of course totally unexpected!

A similar extensive stability analysis was proposed in \cite{Albert:2005} to account for the degeneracy of the two polarization modes of VCSELs. This question is mathematically analogous to our problem invoking two longitudinal modes except that there is no theoretical estimations of the coupling factor in that case. A fairly comprehensive system behavior was reported, including the polarization switching of a VCSEL based on the gain saturation in a manner consistent with what is observed experimentally.

\section{Quantum-dots dual-mode lasers}\label{QDmodel}

QD lasers are significantly different from bulk or QW semiconductor lasers discussed above. Indeed QDs once created have little interaction with each other and substantially the same ability to capture a free carrier from the barrier. Several things differentiate strongly this type of material gain with previous ones:

\begin{enumerate}
\item The total number of active QD in resonance with a given mode is determined by the construction, and even if this number is high, it remains much smaller than the possible number of states allowed in the conduction band of a bulk or QW gain medium.
\item QDs are much less interconnected than the excited states in the conduction band of a more usual semiconductor laser. In particular, they cannot directly exchange carriers either within the QD population of the same mode, or between QD populations addressing different modes. In practice the exchange of carriers between two QDs must involve first thermionic emission to the barrier and capture of this carrier by another QD. This indirect process considerably weakened the coupling as compared to the case of bulk or QW lasers. Given the values ??of the energy levels at stake in a system with InAs QDs in InP barrier, we neglect this coupling here although it could be taken into account in a subsequent Monte Carlo numerical model of dual-mode operation of the laser.
\item According to the QDs manufacturing method, a monoatomic InAs wetting layer appearing at the growth interface is often common to all dots. Owing to this layer, the QDs are likely to have a kind of direct electrical coupling that can transfer the excitation of a dot family to another family addressing another mode. The coupling effectiveness obviously depends on the average distance between dots, and on the nature and thickness of this wetting layer which is intimately dependent of the growth technology.
\item The homogeneous optical linewidth of a single QD is estimated between 3 and 12 meV at room temperature \cite{Gammon:1996, Borri:1999, Matsuda:2001, Kammerer:2002, Bafna:2006}. A direct optical coupling thus necessarily occurs between two QD populations addressing optical modes separated energetically by $\approx 4.2$~meV to obtain a beating at $\approx 1$~THz.
\end{enumerate}

We will  evaluate these situations in sequence and estimate analytically how the various coupling effects interact with the stability of the dual-mode laser.

\subsection{Uncoupled quantum-dots}\label{uncoupled}

Let us first assume an ideal dual-mode QD laser with two uncoupled dots families perfectly centered on the two optical modes allowed by the dual-mode cavity. Rate-equations thus write
\begin{subequations}
\label{rateQD}
\begin{align}
\frac{d N_i}{d t} &= \alpha_i \kappa - \frac{N_i}{{\tau_n}_i} - A_i \left( N_i (S_i+1)-P_i S_i \right) \\
\frac{d S_i}{d t} &= -\frac{S_i}{{\tau_p}_i}+A_i \left( N_i (S_i+1)-P_i S_i \right) 
\end{align}
\end{subequations}
with $i = 1,2$ and $B_i = N_i +  P_i$ the total number of QD addressing mode $i$, $N_i$ the number of  excited dots and $P_i$ the number of unexcited dots, $\kappa$ the total pumping and $0 \le \alpha_i \le 1$ with $\alpha_1+\alpha_2=1$ two parameters that reflect the distribution of the pump between the two QD families. As above $A_i$ terms, ${\tau_n}_i$ and ${\tau_p}_i$ account for the modal gains, the carrier lifetime in the excited state and the photon lifetime within the cavity for each of the two modes.

The reality of a QD laser is not exactly the one described here. In practice, the size dispersion of quantum dots produces an inhomogenously broadened gain  covering almost uniformly the two allowed modes with respect of our assumption of a THz-beating dual-mode laser. As a result many dots are excited but unable to feed a laser mode, thereby contributing to an increase of the laser temperature and a degradation of its quantum efficiency. Although neglected in Eqs.~\eqref{rateQD}, this effect could be taken into account  by an adequate reduction of the effective pumping term $\kappa$ for both QD populations.

A careful inspection of Eqs.~\eqref{rateQD} clearly shows that there is no direct coupling between the two QD populations. The stationary solution obtained by canceling the derivatives and keeping only the dual-mode physical solution with positive populations is
\begin{subequations}
\label{rateQDstat}
\begin{align}
N_i^\mathrm{ss}&= \frac{1}{4 {\tau_p}_i} \left( \tau_i +\tau'_i - \sqrt{\left( \tau_i - \tau'_i \right)^2 + 4 {\tau_n}_i \tau'_i} \right)\\
S_i^\mathrm{ss}&= \frac{1}{4 {\tau_n}_i} \left( \tau'_i -\tau_i + \sqrt{\left( \tau_i - \tau'_i \right)^2 + 4 {\tau_n}_i \tau'_i} \right)
\end{align}
\end{subequations}
where $i=1,2$ depending on the considered mode and $\tau_i=\frac{1}{A_i}+B_i {\tau_p}_i+ {\tau_n}_i$ and $\tau'_i=2 \alpha_i \kappa {\tau_p}_i {\tau_n}_i$. Of course the number of unexcited QDs directly derived, $P_i^\mathrm{ss}=B_i-N_i^\mathrm{ss}$. Similarly the threshold is calculated analytically as the pumping level where a break occurs in the slope of the photon stationary solution, $\tau'_i=\tau_i-2{\tau_n}_i$.\footnote{Note that if we had neglected the spontaneous emission, \emph{i.e.} the term `$+1$' in Eqs.~\eqref{rateQD}, the threshold would have been $\tau'_i=\tau_i$ or ${\kappa_\mathrm{th}}_i=\frac{1}{2 \alpha_i {\tau_n}_i} \left( B_i + \frac{1}{A_i {\tau_p}_i} \right)$.} In fact, Eqs.~\eqref{rateQDstat} describe a behavior of two completely independent lasers with quantum efficiencies and thresholds  evidencing no modal competition.
Assuming $S_i^\mathrm{ss} \gg 1$, it is easy to verify with the method proposed by Lamb that this solution is stable and that it can be generalized to any number of $n$ juxtaposed modes.

\subsection{Quantum-dots electrically coupled through the wetting layer}\label{wet}

To go beyond this first simple view of a QD laser  where the decoupling between  modes is total and therefore where no modal competition will affect the two-mode regime, we now imagine that the two QD populations  can interact with each other directly. Presumably an excited dot of the first family can transfer to an unexcited dot of the second family, the difference in photon energy being supplied or absorbed by the states involved in the wetting layer. This direct interaction is similar to a chemical-like equilibrium law between populations
\begin{equation}\label{actionmasse}
N_1 +P_2 \leftrightarrow N_2+P_1
\end{equation}

In practice Eqs.~\eqref{rateQD} becomes 
\begin{subequations}
\label{rateQDlam}
\begin{align}
\begin{split}
\frac{d N_1}{d t} &= \alpha_1 \kappa + k_1 N_2 P_1 - k_2 N_1 P_2 - \frac{N_1}{{\tau_n}_1} \\
& \quad - A_1 \left( N_1 (S_1+1)-P_1 S_1 \right)
\end{split} \\
\begin{split}
\frac{d N_2}{d t} &= \alpha_2 \kappa +  k_2 N_1 P_2 - k_1 N_2 P_1 - \frac{N_2}{{\tau_n}_2} \\
& \quad - A_2 \left( N_2 (S_2+1)-P_2 S_2 \right)
\end{split} \\
\frac{d S_1}{d t} &= -\frac{S_1}{{\tau_p}_1}+A_1 \left( N_1 (S_1+1)-P_1 S_1 \right) \\
\frac{d S_2}{d t} &= -\frac{S_2}{{\tau_p}_2} +A_2 \left( N_2 (S_2+1)-P_2 S_2 \right)
\end{align}
\end{subequations}
where $k_i$ are constants reflecting the exchange rate of the excited states between the two QD families. Clearly if $k_i = 0$ QDs are not coupled anymore and one retrieve the previous situation. Instead, if $k_i$ constants are strong enough their corresponding rates in Eqs.~\eqref{rateQDlam} quickly become of paramount importance as compared to any other rate. This is because they are assigned to products of two carrier populations that correspond to the highest values involved in Eqs.~\eqref{rateQDlam}. Thus the proper operation of the laser is obtained when these two rates almost annihilate or equivalently
\begin{equation*}
k_2 N_1 P_2 \approx k_1 N_2 P_1 \; \Longleftrightarrow \; N_1 P_2 \approx C N_2 P_1 \quad \text{with} \quad C=\frac{k_1}{k_2}
\end{equation*}
which is the analogue of chemical equilibrium described by Eq.~\eqref{actionmasse} where carrier populations are strongly coupled.

For the sake of simplicity we now neglect the terms in this `$+1$' that account for spontaneous emission in each mode in Eqs.~\eqref{rateQDlam}. Dual-mode stationary solution for carriers is then simply given by
\begin{equation}
N_i^\mathrm{ss}=\frac{B_i}{2} + \frac{1}{2 A_i {\tau_p}_i} \quad \text{with $i=1,2$}
\end{equation}
But the dual-mode stationary solution for photons is much more complicated in the general case. Two special cases are however interesting to consider:
\begin{itemize}
\item[(a)] when the coupling induced by the wetting layer are identical, $k_i=k$
\begin{multline}
S_i^\mathrm{ss}=\alpha_i \kappa {\tau_p}_i \\
- \frac{1}{2} \left(  \frac{B_i {\tau_p}_i}{{\tau_n}_i} + \frac{1}{A_i {\tau_n}_i} + k {\tau_p}_i \left[ \frac{B_j}{A_i {\tau_p}_i} - \frac{B_i}{A_j {\tau_p}_j} \right]  \right) \label{Ska}
\end{multline}

\item[(b)] when the material and cavity parameters are identical for the two modes, \emph{i.e.} $A_i=A$, $B_i=B$, ${\tau_p}_i=\tau_p$, ${\tau_n}_i=\tau_n$.
\begin{multline}
S_i^\mathrm{ss}=\alpha_i \kappa {\tau_p} \\
- \frac{1}{2} \left( \frac{B {\tau_p}}{{\tau_n}} + \frac{1}{A {\tau_n}} +(k_i-k_j) \tau_p \left[ \frac{1}{2 A^2 \tau_p^2} - \frac{B^2}{2} \right]  \right) \label{Skb}
\end{multline}

\end{itemize}
where $(i,j)=(1,2)$ or $(2,1)$.
If cases (a) and (b) occur simultaneously, the situation is similar to that of \S\ref{uncoupled}  because the $k_i$-dependent terms cancel in Eqs.~\eqref{Ska} and \eqref{Skb}, therefore the coupling by the wetting layer becomes inefficient. Out of this particular case, the coupling by the wetting layer induces a differentiation between the thresholds of the two laser modes obtained approximately here by setting $S_i=0$ in \eqref{Ska} et \eqref{Skb}. The mode that had the lowest threshold before coupling feeds its population of excited QDs by direct transfer from the one who had the highest threshold, the result is an amplification of the threshold difference. In the extreme case where the constants $k_i$ are larges, the population of excited QDs which reaches the first its stimulated emission threshold is clamped and in turn saturates the population of excited QDs for the other mode to a level below its own threshold. Strong coupling between QD populations thus leads to the destruction of the two-mode regime.

In the case where the coupling is sufficiently moderate so that the dual-mode laser regime is preserved, the stability analysis is conducted as in \S\ref{unic}. We report the stationary values plus a small fluctuation in the rate equations \eqref{rateQDlam} to obtain the evolution matrix of these fluctuations
\begin{equation}
\mathbf{\Theta} = \begin{pmatrix}
-\sigma_1&\zeta_2&-\delta_1&0\\
\zeta_1&-\sigma_2&0&-\delta_2\\
\beta_1&0&0&0\\
0&\beta_2&0&0
\end{pmatrix}
\end{equation}
with
\begin{align*}
\beta_1&=2 A_1 S_1^\mathrm{ss} &
\beta_2&=2 A_2 S_2^\mathrm{ss} \\
\zeta_1&=k_1 N_2^\mathrm{ss} + k_2 P_2^\mathrm{ss} &
\zeta_2&=k_2 N_1^\mathrm{ss} + k_1 P_1^\mathrm{ss} \\
\delta_1&=A_1(N_1^\mathrm{ss}-P_1^\mathrm{ss}) &
\delta_2&=A_2(N_2^\mathrm{ss}-P_2^\mathrm{ss}) \\
\sigma_1&=\beta_1+\zeta_1+1/{\tau_n}_1 &
\sigma_2&=\beta_2+\zeta_2+1/{\tau_n}_2 
\end{align*}
which are all positive terms since the population inversion above the threshold requires $N_i^\mathrm{ss}>P_i^\mathrm{ss}$. 

Fluctuations may return to equilibrium if and only if the eigenvalues $\lambda$ of $\mathbf{\Theta}$ have negative real part. These eigenvalues are the roots of the characteristic polynomial
\begin{multline}
\mathcal{P}(\lambda)=\lambda^4+(\sigma_1+\sigma_2)\lambda^3+(\pi_1+\pi_2+\sigma_1\sigma_2-\zeta_1\zeta_2)\lambda^2 \\
+(\pi_1\sigma_2+\pi_2\sigma_1)\lambda+\pi_1\pi_2.
\end{multline}

Because $\sigma_i>\zeta_i$, all the coefficients of $\mathcal{P}(\lambda)$ are positive. According to the Routh-Hurwitz criterion, all roots will then have their real part negative if and only if \cite{Hurwitz:1964}
\begin{multline}\label{inequal1}
(\sigma_1+\sigma_2)(\pi_1+\pi_2+\sigma_1\sigma_2-\zeta_1\zeta_2) \\
- (\pi_1\sigma_2+\pi_2\sigma_1) 
>\frac{(\sigma_1+\sigma_2)^2\pi_1\pi_2}{\pi_1\sigma_2+\pi_2\sigma_1}
\end{multline}
We can rewrite the difference of the two members of \eqref{inequal1}
\begin{multline}\label{hurwitz}
(\sigma_1+\sigma_2)(\sigma_1\sigma_2-\zeta_1\zeta_2)+
(\pi_1\sigma_1+\pi_2\sigma_2)-\frac{(\sigma_1+\sigma_2)^2\pi_1\pi_2}{\pi_1\sigma_2+\pi_2\sigma_1}\\=
(\sigma_1+\sigma_2)(\sigma_1\sigma_2-\zeta_1\zeta_2)+\sigma_1\sigma_2
\frac{(\pi_1-\pi_2)^2}
{\pi_1\sigma_2+\pi_2\sigma_1} 
\end{multline}
and remark that since $\sigma_i>\zeta_i$ this difference is always positive.
We then conclude that the dual-mode stationary solution obtained from Eqs.~\eqref{rateQDlam} is stable. It proves that the coupling of carriers by a chemical-like equilibrium law between the two QD populations does not prohibit a dual-mode behavior in the sense of modal competition, even if it is likely to push the threshold up to an unacceptable value for one of the lasing mode if too strong.

\subsection{Quantum-dots optically coupled through the homogeneous linewidth}

We now consider the two QD families coupled only through the homogeneous gain width of each dot, the rate equations then become
\begin{subequations}
\label{rateQDhom}
\begin{align}
\frac{d N_1}{d t} &= \alpha_1 \kappa - \frac{N_1}{{\tau_n}_1} - A_1( N_1 - P_1) (S_1 + \epsilon S_2) \\
\frac{d N_2}{d t} &= \alpha_2 \kappa - \frac{N_2}{{\tau_n}_2} - A_2 (N_2 - P_2) (S_2 + \epsilon S_1) \\
\frac{d S_1}{d t} &= -\frac{S_1}{{\tau_p}_1} + A_1 (N_1 -P_1) S_1 + \epsilon A_2 (N_2 - P_2) S_1 \\
\frac{d S_2}{d t} &= -\frac{S_2}{{\tau_p}_2} +A_2 (N_2 - P_2) S_2 + \epsilon A_1(N_1 - P_1) S_2
\end{align}
\end{subequations}
with of course the particle conservation of each QD family $N_i+P_i=B_i$ and assuming above  threshold operation for each mode which removes the spontaneous emission term `$+1$' in photon density equations. The variable $\epsilon$ here accounts for the direct optical coupling between dots of the two modes. It depends on the value of the homogeneous gain width and thus ranges from $\approx 10$\% to $60$\% for a 1~THz frequency separation, depending on whether one considers a  broadening of 3~meV or 10~meV.

In the general case the expression of the stationary solutions is complex, but in the particular case of two populations having the same QD material parameters ${\tau_n}_i={\tau_n}$, ${\tau_p}_i={\tau_p}$, $A_i=A$, $B_i=B$, and  identical pumps $\alpha_i=1/2$, the stationary solutions are
\begin{subequations}
\label{stathom}
\begin{align}
N_i^\mathrm{ss}&=\frac{B}{2} + \frac{1}{2 A (1+\epsilon)\tau_p} \\
S_i^\mathrm{ss}&=\frac{\kappa \tau_p}{2}  -\frac{B \tau_p}{2 \tau_n} - \frac{1}{2 A ( 1+ \epsilon ) \tau_n} \label{sss}
\end{align}
\end{subequations}
These formulas are close to those obtained in  \S\ref{uncoupled} for uncoupled QDs. In fact only the term $1+\epsilon$ differs and makes the calculated thresholds  apparently lower in this case as compared to uncoupled QDs. This distortion effect does not hold if we consider that the differential gain of a laser with high  homogeneous  linewidth will inevitably be far lower than that of a laser with a narrower linewidth. No decrease in thresholds related to modal coupling should be expected as a simplistic reading of Eq.~\eqref{sss} suggests.

Like in previous section with carriers coupled by the wetting layer, the stability analysis of the dual-mode regime has been conducted in the general case. Again this stability is governed by the sign of the real parts of the eigenvalues of the evolution matrix
\begin{equation}
\mathbf{\Theta} = \begin{pmatrix}
- \sigma_1 & 0 & - \delta_1 &- \epsilon \delta_1\\
0 & - \sigma_2 & - \epsilon \delta_2 & - \delta_2\\
\beta_1 & \xi_2 & 0 & 0\\
\xi_1 & \beta_2 & 0 & 0
\end{pmatrix}
\end{equation}
Again we have chosen to define as positive all the terms of this matrix
\begin{align*}
\beta_1&=2 A_1 S_1^\mathrm{ss}  &
\beta_2&=2 A_2 S_2^\mathrm{ss} \\
\xi_1&=2 \epsilon A_1 S_2^\mathrm{ss} &
\xi_2&=2 \epsilon A_2 S_1^\mathrm{ss} \\
\delta_1&=A_1(N_1^\mathrm{ss}-P_1^\mathrm{ss}) &
\delta_2&=A_2(N_2^\mathrm{ss}-P_2^\mathrm{ss}) \\
\sigma_1&=\beta_1 + \xi_1 + 1/{\tau_n}_1 &
\sigma_2&=\beta_2 + \xi_2 + 1/{\tau_n}_2
\end{align*}
and the eigenvalues are the roots of the following characteristic polynomial (assuming $\pi_i=\delta_1(\epsilon\xi_i+\beta_i)$ for further simplification)
\begin{multline}\label{eigenIII2}
\mathcal{P}(\lambda)=
\lambda^4+(\sigma_1+\sigma_2)\lambda^3+(\sigma_1\sigma_2+\pi_1+\pi_2)\lambda^2 \\
+(\pi_1\sigma_2+\pi_2\sigma_1)\lambda+(1-\epsilon^2)\delta_1\delta_2(\beta_1\beta_2-\xi_1\xi_2)
\end{multline}

As $\xi_1\xi_2=\epsilon^2\beta_1\beta_2$,  the constant term can be written more simply as $c=(1-\epsilon^2)^2\delta_1\delta_2\beta_1\beta_2$, and all the coefficients of $\mathcal{P}(\lambda)$ are then positive since $0<\epsilon<1$.
According to the Routh-Hurwitz criterion, $\mathcal{P}(\lambda)$ roots exhibit negative real parts if and only if the following amount $r$ is positive \cite{Hurwitz:1964}
\begin{equation*}
r=(\sigma_1+\sigma_2)(\pi_1+\pi_2+\sigma_1\sigma_2)
- (\pi_1\sigma_2+\pi_2\sigma_1)-\frac{(\sigma_1+\sigma_2)^2c}{\pi_1\sigma_2+\pi_2\sigma_1}
\end{equation*}
yielding after simplification
\begin{equation}
r=(\sigma_1+\sigma_2)\sigma_1\sigma_2+
(\pi_1\sigma_1+\pi_2\sigma_2)-\frac{(\sigma_1+\sigma_2)^2c}{\pi_1\sigma_2+\pi_2\sigma_1}
\end{equation}
where we see that $r$ is an increasing function of $\epsilon$ on $[0,1]$ as a sum of functions of such type. The required proof is thus obtained by just checking if the value at $\epsilon=0$ is positive. Noting that at $\epsilon=0$, $c=\pi_1\pi_2$, allows to write as in \S\ref{wet} Eq.~\eqref{hurwitz} 
\begin{equation*}
(\pi_1\sigma_1+\pi_2\sigma_2)-\frac{(\sigma_1+\sigma_2)^2c}{\pi_1\sigma_2+\pi_2\sigma_1}=
\sigma_1\sigma_2
\frac{(\pi_1-\pi_2)^2}
{\pi_1\sigma_2+\pi_2\sigma_1}
\end{equation*}
This proves that $r>0$ and concludes again to the stability of the dual-mode regime in that case. It should be noted, however, that if $\epsilon$ tends to $1$, which corresponds to a gain width very large compared to the energy gap of the two modes, then two eigenvalues tend to zero making the behavior of the dual-mode laser extremely unstable. This is a limit case where the laser is very to close to a bistable behavior but this is not the scenario we have chosen for THz radiation by photomixing with an energy gap of 4.1~meV between modes and QD homogeneous linewidths between 3 and 10~meV.

\section{Conclusion}\label{conclusion}

We studied different semiconductor laser types devoted to operate continuously in stable dual-mode emission. The proposed analysis is based on the rate-equations and Lamb's theory. After considering the case of laser with bulk or QW gain mediums for which it is demonstrated that the natural behavior is bistable and not dual-mode, we evaluated analytically what should be expected from QD semiconductor lasers. We have considered uncoupled as well as coupled QDs either electrically through a direct exchange of excitation by the wetting layer, or optically through the homogeneous broadening of the gain. In all cases we have shown analytically that a stable dual-mode emission is possible.

If the objective of building a dual-mode semiconductor laser producing a stable beating frequency in the THz range is highlighted, the selection of a QDs as gain medium is the most suitable as compared to bulk or QWs that inevitably lead to a bistable behavior. The incorporation of QDs in the active membranes of future photonic crystal optoelectronic components for the 2.5D THz radiation by photomixing is on the way \cite{Kusiaku:2012}.

\begin{acknowledgments}
This work was supported by the french Agence Nationale de la Recherche under contracts ANR-08-NANO-052 BASTET and ANR-12-MONU-19 MARMOTE.
\end{acknowledgments}

%

\end{document}